\documentstyle[prd,aps]{revtex}
\input epsf.tex
\def\DESepsf(#1 width #2){\epsfxsize=#2 \epsfbox{#1}}
\begin{document}
\draft
\twocolumn[\hsize\textwidth\columnwidth\hsize\csname @twocolumnfalse\endcsname
\preprint{ UMD-PP-97-108 OITS-627 OSURN-325 {}~hep-ph yymmnn}
\title{Explaining the HERA Anomaly Without Giving Up R-parity Conservation}
\author{B. Dutta$^{1}$, R. N. Mohapatra$^{2}$ and S. Nandi$^{3}$}
\address{$^{(1)}${\it Institute of Theoretical Science, University of Oregon,
Eugene, OR 97403}}
\address{$^{(2)}${ Department of Physics, University of Maryland,  College Park,
Md-20742, USA.}}
\address{$^{(3)}${ Department of Physics, Oklahoma State University, Stillwater,
OK 74078}}
\date{April, 1997}
\maketitle
\begin{abstract}

We point out that in extended supersymmetric models such as supersymmetric
left-right models, it is possible to have leptoquarks that explain the HERA high
$Q^2$ anomaly without giving up R-parity conservation. The leptoquarks belong to 
vectorlike $(2, 2, \pm\frac{4}{3}, 3 {\rm or} 3^*) $ representations of 
$SU(2)_L\times SU(2)_R\times U(1)_{B-L}\times SU(3)_c$ (denoted by $G_{2213}$).
Considerations of D-terms imply that the only phenomenologically viable model is
the one where the leptoquark couples to positrons and the up quark. Unlike the
R-parity violating scenario, the  leptoquarks are accompanied by new
superpartners,the leptoquarkino  which leads to many intersting signatures in
other collider  experiments. At Tevatron, pair productions of the leptoquarkino
will give rise to dilepton signals very distinct from the top productions. These
models can lead to unification of gauge coupling constants at a scale of around
$10^{10}$ GeV implying that grand unification group is not of the usual $SU(5)$
or
$SO(10)$ types but rather an automatically R-parity conserving 
$SU(5)\times SU(5)$ GUT model recently proposed  by one of the authors (R. N.
M.) which leads to a stable proton.
\end{abstract}   

\pacs{  OITS-627\hskip 1cm UMD-PP-97-108\hskip 1cm  OSURN-325\hskip 1 cm}

 \vskip2pc]
If the high $Q^2$ anomaly observed recently in the $e^+p$ scattering
by the H1\cite{H1} and the ZEUS\cite{ZEUS} collaborations is confirmed by future
data, it will be an extremely interesting signal of new physics beyond the
standard model. A very plausible and widely discussed interpretation of this
anomaly appears to be in terms of new scalar particles capable of coupling to
$e^+u$ or
$e^+d$ of scalar leptoquarks\cite{lq} with mass around  200 GeV. Alternative
interpretations based on contact interactions
\cite{contact} or a second $Z'$\cite{godfrey} have been proposed; but attempts
to construct models that lead to the desired properties seem to run into
theoretical problems.
 
The leptoquark must be a spin zero color triplet particle with electric charge 
$5/3$ or $2/3$. The latter electric charge assignment makes it possible to give
a plausible interpretation of the leptoquark as being the  superpartner of the
up-like quark\cite{squark} of the minimal supersymmetric standard model (MSSM)
provided one includes the R-parity violating couplings of the type $\lambda'
QLd^c$ to the MSSM (where $Q,L$ denote the quark and the lepton
$SU(2)_L$ doublets and $u^c, d^c, e^c$ denote the $SU(2)_L$
singlets\cite{haber}) . There are however very stringent upper limits on several
R-parity violating couplings: for instance, if in addition to the $\lambda'$
term described above, one adds the  allowed $\lambda''u^cd^cd^c$ terms to the
superpotential, then it leads to catastrophic proton decay unless
$\lambda'\lambda''\leq 10^{-24}$
\cite{vissani}. There are also stringent limits on $\lambda'_{111}\leq 10^{-4}$
\cite{mhkk} from neutrinoless double beta decay, which forces the lepto-quark to
be $\tilde{c}$ or $\tilde{t}$ rather than the obvious choice
$\tilde{u}$. Furthermore, within such a framework, the lightest supersymmetric
particle (LSP) is no more stable and therefore, there is no cold dark matter
(CDM) candidate in such theories. In view of the fact providing a natural CDM
candidate was long considered an attractive feature of the supersymmetric
models, it may be worthwhile to consider extensions of the MSSM which
incorporate the leptoquark without  giving up R-parity conservation (and hence
the idea of LSP as a natural CDM candidate). It is the goal of this letter to
report on the results of such a study.

The class of supersymmetric theories where R-parity conservation is automatic
can provide an absolutely stable LSP that can act as the CDM candidate. We will
therefore use them as a typical framework for studying the consequences of
leptoquarks incorporated into such theories while at the same time maintaining
automatic R-parity conservation. Minimal versions of such theories are based on
the gauge group $SU(2)_L
\times U(1)_{I_{3R}}\times U(1)_{B-L}$ or $SU(2)_L\times SU(2)_R\times
U(1)_{B-L}$\cite{moh} with usual assignments for the quarks and leptons. We will
consider the latter left-right symmetric gauge group. The quark doublets
$Q\equiv (u,d)$ and $Q^c\equiv (u^c, d^c)$ transform as doublets under the
$SU(2)_L$ and $SU(2)_R$ groups respectively and similarly the lepton doublets
$L\equiv (\nu, e)$ and $L^c\equiv  (\nu^c, e^c)$ respectively. The $SU(2)_R$
gauge symmetry breaking is achieved via the $B-L$ non-singlet isotriplets
$\Delta^c\equiv (1,3,-2)$ and
$\bar{\Delta}^c\equiv (1,3 +2)$ and their lefthanded counterparts added to
maintain left-right symmetry (the numbers in the parenthesis denote the
$SU(2)_L,SU(2)_R$ and $U(1)_{B-L}$ quantum numbers). The standard model symmetry
is broken by the bi-doublet fields $\phi (2,2,0)$.

We augment this model by including the leptoquark fields.  Demanding that the
lepto-quarks couple to $e^+d$ or $e^+u$  leads to the conclusion that they must
belong to the multiplet $(2,2, 4/3,3)$ (denoted $\Sigma_{QL^c}$) and
$(2,2,-4/3, 3^*)$ (denoted by $\bar{\Sigma}_{Q^cL}$). Each of these multiplets
have four scalar fields and four fermion fields which will be denoted in what
follows by the obvious subscript corresponding to their couplings. We denote the
four scalar leptoquarks as
$\bar{\Sigma}_{ue^c}$, $\bar{\Sigma}_{u\nu^c}$, $\bar{\Sigma}_{de^c}$ and
$\bar{\Sigma}_{d\nu^c}$ and their fermionic partners (to be denoted by a tilde
on the corresponding scalar field). Writing down the most general
superpotential, one can easily convince oneself that the resulting theory
maintains the property of automatic R-parity conservation. Thus the lightest
neutralino LSP will be stable in this model and can serve as the cold dark
matter.
   
Before discussing the application of this model to explain the HERA anomaly, let
us first discuss the mass spectrum of the model. The superpotential for this
model will have a direct mass term of the form
$M_0\Sigma \bar{\Sigma}$ which will imply that the fermionic fields in the
leptoquark multiplet will have a masses $M_0$ prior to symmetry breaking. There
may be other contributions to these masses from radiative corrections which will
split the degeneracy implied by the above mass term.

As far as the scalar leptoquark states are concerned, their masses will receive
several contributions:  first a direct common contribution from the $M_0$ term
given above. After symmetry breaking, the D-terms of the various gauge groups
will contribute. There is also soft SUSY breaking contribution along with the
radiative correction. Assuming that the
$SU(2)_R$ symmetry is broken by the vev's $<\Delta^c>=v_1$,
$<\bar{\Delta}^c>=v_2$ and the
$SU(2)_L$ symmetry is broken by the two $\phi$ vev's as $diag <\phi_u>=
(0,vsin\beta )$ and $diag<\phi_d>=(vcos\beta,0)$,  we can write the masses for
the various scalar leptoquarks as follows:
\begin{eqnarray} M^2_{\Sigma_a}&=& M^2_0 +
(I^a_Rg^2_{2R}-\frac{B-L}{8}g^2_{B-L})(v^2_1-v^2_2)
\\\nonumber&+&\frac{g^2_{2L}v^2I_{3L}}{4} cos 2\beta +\Delta_m^2+{\rm Radiative\,
correction}
\end{eqnarray}$\Delta_m^2$ is the soft SUSY breaking contribution. The values of
the $I^a_R$ and
$B-L$ are given in Table I:

\begin{table}
\begin{tabular}{|c||c||c||c|} \hline states & $I^a_R$ & $B-L$ & $I_{3L}$ \\
\hline
$u^ce$ & $\frac{1}{2}$ & $\frac{4}{3}$ & $\frac{1}{2}$ \\\hline
$u^c\nu$ & $\frac{1}{2}$& $\frac{4}{3}$& $-\frac{1}{2}$ \\\hline
$ue^c$ & $-\frac{1}{2}$ &$-\frac{4}{3}$ & $-\frac{1}{2}$ \\\hline
$de^c$ & $-\frac{1}{2} $ & $-\frac{4}{3}$ & $\frac{1}{2}$ \\\hline
$d^c\nu$ & $-\frac{1}{2}$ & $\frac{4}{3}$ & $-\frac{1}{2}$\\\hline
$d^ce$ & $-\frac{1}{2}$ & $\frac{4}{3}$ & $-\frac{1}{2}$\\ \hline
\end{tabular}
\caption{The $I_{3R}$ and $B_L$ quantum numbers of the various leptoquark states
}
\end{table}
 The radiative correction can  be positive or negative. From table I and Eq.(1),
we see easily that if $v_1 < v_2$, then the lightest leptoquark state is
the first entry in the table I which corresponds to the conjugate of 
$\bar{\Sigma}_{ue^c}$ since $g^2_{2R}> 2g^2_{B-L}$ in the left-right models.
Furthermore, interestingly enough for this choice of vev's, assuming the
combined  $\Delta_m^2+{\rm Radiative\, correction}$ to be smaller than the the
D-term, the leptoquarkino states are heavier than the lightest leptoquark
state.  We assume their masses to be in the range of 300 to 400 GeV.  In a
subsequent section we will obtain a lower bound on the leptoquarkino mass from
the present collider data.

Turning to the couplings of the leptoquarks $\Sigma $ and $\bar{\Sigma}$ to
quarks and leptons, it is given by the superpotential:
\begin{eqnarray} W_{lq}= \lambda_{ij} (\Sigma Q^c_iL_j +\bar{\Sigma} Q_i L^c_j)
\end{eqnarray} (where $i,j$ are generation indices). Let us assume for
simplicity that 
$\lambda_{ij}$ are diagonal in the mass eigenstate basis for the quarks and
leptons. Explanation of the HERA high $Q^2$ anomaly seems to require
$\lambda_{11}\simeq 0.05$ which we will assume from now on and mass of the
scalar field in $\Sigma$ (assumed to be $\Sigma_{u^ce}$ from the above mass
arguments) around 200 GeV.  As far as the other couplings go, they will be
strongly constrained by the present experimental upper limits on the low energy
processes  such as $\mu\to e\gamma$, 
$\tau\to e\gamma$ etc. which will arise at the one loop level from the exchange
of $\Sigma$ and $\bar{\Sigma}$. There are also tree level diagrams which can
lead to rare processes such as $K\to \pi e^-\mu^+$. These  processes imply an
upper limit of $\lambda_{22}\leq 2\times 10^{-3}$. The most stringent upper
limit on $\lambda_{22}$ comes from the upper limit on the process $K^0_L\to
\mu^+e^-$ and yields $\lambda_{22}\leq 2\times 10^{-4.5}$. Turning to
$\lambda_{33}$, the most stringent limits arise from the present upper limit on
the branching ratio for the process $\tau\to e\gamma$  which the 1996 Particle
data tables give as $\leq 1.1\times 10^{-4}$\cite{pdg}. This implies a weak
upper limit on $\lambda_{33}\leq 0.2$. Thus the third generation leptoquark
coupling $\lambda_{33}$ could in principle be comparable to $\lambda_{11}$. If
the efficiency for the detection of $\tau$ leptons at HERA were comparable to
the detection efficiency for electrons, that could also severly limit
$\lambda_{33}$.  Finally we also note that such a value for $\lambda_{11}$ is
also consistent with the present data on the parity violation in atomic physics.
Thus it appears that our leptoquark couplings are consistent with all known low
energy data. It is clear from the above discussion that the leptoquarks do not
respect $\mu-e$ universality.
                                             
The existence and properties of the leptoquarkino provides a new and  unique
signature for our model as compared to all other proposals to  explain the HERA
anomaly. To see this note that in hadron colliders, we can produce pairs of
leptoquarkinos at the same rate as the $t\bar{t}$ pair due to identical color
content. Moreover, due to R-parity conservation the leptoquarkino decay leads to
a missing energy signal as follows:
$\tilde{\Sigma}\to e^+\tilde{u}$ with $\tilde{u}\to u + \chi^0$ or
$\tilde{\Sigma}\to u+\tilde{e^+}$ with $\tilde{e^+}\to e^+ + \chi^0$. In both
the cases we have $e^+,u$ plus missing energy in the final state. If the
$\lambda_{33}$ coupling is comparable to $\lambda_{11}$ as is allowed, then the
branching ratio for leptoquarkino decay to electrons  will be 50\%. This signal
is similar to the top signal at the Tevatron with one crucial difference.  The
dilepton branching ratio from the leptoquark pairs is 100$\%$ compared to only
10$\%$ from the top pairs. There will be no $\mu^{+} \mu^{-}$ events. The
branching ratio to $e^{+} e^{-}$ will be 25$\%$ compared to only 1$\%$ from the
top. The present observations should therefore lead to lower limits on the mass
of the lightest leptoquarkino pairs. Thus an excess of dilepton pairs over that
expected from the top productions, or an excess in $e^{+} e^{-}$ channel over
$\mu^{+} \mu^{-}$ will be a clear signal of leptoquark productions at Tevatron.
In addition, leptoquarks will also give rise to harder  leptons and larger
missing energy events. In Fig. 1, we plot the cross section for the
leptoquarkino pair production. We see that for the present combined Tevatron
data of about 200 pb$^{-1}$, there would be 6 events of type $e^+e^-$jj plus
missing energy for
$M_{\tilde{\Sigma}_{u^ce}} = 250$ GeV, assuming a detection efficiency of 0.2
for the $e^+e^-$ mode. At present CDF collaboration has only one such
event\cite{cdf}, and the D0 collaboration has also 1 event \cite{D0}. Using
these two events, and the the cross sections given in Fig. 1, we obtain a lower
bound on the leptoquarkino mass of about 290 GeV ( assuming the detection
efficiency to remain  0.2 for the higher masses). Any evidence for an excess of
dielectron events would be interpretable interms of this new leptoquarkino. We
urge the CDF and D0 coluboration to look for such excess, and also to look for
any dilepton event in which the lepton  $P_T$ or missing energy do not fit the top
productions. The recent CDF analysis \cite{talk} shows that the leptoquark mass
bound at 95$\%$ CL is 210 GeV assuming the branching ratio is 1. In this model
however the branching ratio is less than 1, since the leptoquark can also decay into
the top quark and the $\tau$ lepton.

So far we have discussed the signals of leptoquarkinos in the usual SUSY
theories. If however the supersymmetry breaking is communicated to the
obseravable sector by the SM gauge ineteraction \cite{DN}, the signal is
different. Let us first consider the case when neutralino is the NLSP in these
models. The  leptoquarkino decays into a spositron and u quark (squarks masses
are large in these models). The spositron decays (100$\%$) into a positron and a
neutralino and the neutralino then decays (100$\%$) into a hard photon and a
gravitino. Consequently the final state in the leptoquarkino pair production
process has 
$e^+e^-\gamma\gamma$jj plus missing energy without any SM background.  If
lighter stau is the NLSP and neutralino is the NNLSP, spositron decays (100$\%$)
into a positron and a neutralino and the neutralino then decays (100$\%$) into a
$\tau$ and a lighter stau. The lighter stau then decays into a tau and a
gravitino. Consequently the  final state in the leptoquarkino pair production
process in the Tevatron has 
$e^+e^-2\tau^+2\tau^-$jj plus missing energy. Out of these six leptons in the
final state, one
$\tau$ pair (produced from the decay of stau) has much higher $P_T$ than the
other leptons. 

 One could also look for the signals of leptoquarkino production in
$e^+\gamma$ colliders. The leptoquark or leptoquarkino can be singly produced in
such a  collider. The scalar leptoquark will be produced in the association of 
an antiquark\cite{don}. As discussed before in the usual SUSY theories, the
leptoquark further decays  into a quark and a positron with  the final state
consisting of positron +jets. The leptoquarkino however will be
 produced along with a squark.  This leptoquarkino will then decay into a 
lepton and a squark. The final state has electron +jets+missingenergy.  The
missing energy part will then disentangle the leptoquarkino signal. In the gauge
mediated SUSY breaking scenario, the final state has either a hard photon or
$\tau^+\tau^-$ along with electron +jets+missing energy.

In $e^{+}e^{-}$ collider the signal for this leptoquarkino would be  jets
+missing energy, since a squark anti-squark pair will be produced from a  t
channel exchange of
 leptoquarkino. The exchange of a t channel leptoquark however   gives rise to
jets without any missing energy.  It is not possible for LEPII to see the sign
al of leptoquarkino due to the large squark mass. A higher energy machine is 
needed for that purpose. A detailed study of the discovery prospects of the
leptoquarkino in different  colliders will be presented elsewhere.

Another important implication of our model is that there does not seem to any
operator which can give a non-negligible charged current signal at HERA. Thus
observation of charged current like events above background at HERA will be an
evidence against this model.
 
Let us now discuss the possible implications of the existence of a low mass
leptoquark supermultiplet for unification. We will assume that the physics
immediately beyond MSSM (i.e. in the TeV region) is described by a
supersymmetric left-right symmetric model (SUSYLR) with the addition of the
leptoquark multiplets described above. The evolution of the gauge couplings 
depend not only on the individual beta functions but also on the nature of the
final unification group which determines the normalizations of the  various
gauge couplings at low energies. For instance, if we envision embedding the
SUSYLR group within a simple group such as $SO(10)$ or $SU(16)$ etc, the
properly normalized weak hypercharge, $I_Y$ is given by the familiar formula $
I_Y = \sqrt {\frac{3}{5}}(\frac{Y}{2})$ and the properly normalized $B-L$ charge
$I_{BL}$ is given by $I_{BL} = \sqrt{\frac{3}{2}} (\frac{B-L}{2})$. This leads
to the matching formula for the weak hypercharge coupling to be
$\alpha^{-1}_{I_Y}=\frac{3}{5}\alpha^{-1}_{2R}+
\frac{2}{5}\alpha^{-1}_{BL}$ at $M_R$.  Evolving our model with this kind of
unification leads to a unification scale of about $10^{9.5}$ GeV assuming that
at the $W_R$ scale, we have two bidoublets, right handed triplets of type $(1,
3, -2, 1)+(1,3, +2,1)$ without their left-handed partner in addition to the
aforementioned leptoquark multiplets. This is therefore unacceptable since it
will lead to catastrophic proton decay via the exchange of gauge bosons that
violate baryon number in the $SO(10)$ models. We therefore consider a different
kind of embedding of SUSYLR into an
$SU(5)\times SU(5)$ model suggested in a recent paper\cite{RNM}.

\vskip2pc

\noindent{\it $SU(5)\times SU(5)$ embedding of SUSYLR with leptoquarks}

\vskip2pc

First let us briefly recapitulate the fermion assignments of the model: they
belong to the $(\bar{\bf 5}+{\bf 10}, 1)+(1, {\bf 5}+{\bar{\bf 10}})$
multiplet\cite{RNM} for every generation. One therefore needs a weak singlet
vectorlike pair of $D, U$ quarks and a heavy weak singlet vectorlike charged
lepton $E^{\pm}$. The lefthanded $(\bar{\bf 5}, 1)$ multiplet then consists of
$(D^c_1, D^c_2, D^c_3, e^-, \nu_e)$ whereas the righthanded multiplet $(1, {\bf
5})$ is given by $(D_1,D_2,D_3, e^c,\nu^c)$. The assignments to the {\bf 10}
dimensional representations are easily obtained\cite{RNM}. For instance, the
fermion assignment in $({\bf 10},1)$  representation is given by
\begin{eqnarray}
\left(\begin{array}{ccccc} 0 & U^c_3 & -U^c_2 & u_1 & d_1 \\ -U^c_3 & 0 & U^c_1
& u_2 & d_2 \\ U^c_2 & -U^c_1 & 0 & u_3 & d_3 \\ -u_1 & -u_2 & -u_3 & 0 & E^+ \\
-d_1 & -d_2 & -d_3 & -E^+ & 0 \end{array}\right)
\end{eqnarray} and similarly for the $(1,{\bf  \bar{10}})$ multiplet which
contains the other chirality states for the above fields arranged exactly the
same way. The Higgs mechanism of the model is implemented by multiplets of type
$({\bf 5, \bar{5}})+ ({\bf \bar{5}, 5})$ and they lead to  the vectorlike quarks
and leptons acquiring mass at the unification scale. This leaves the low energy
theory to be the usual SUSYLR model. The SUSYLR symmetry is broken down to MSSM
by the Higgs multiplets of type $(1, {\bf 15})+(\bar{\bf 15}, 1)$ which also
leads to the see-saw mechanism for neutrino masses. Below the $W_R$ scale the
theory is MSSM with the important  difference that now the R-parity violating
couplings are automatically absent. In order to accomodate the leptoquarks in
this model, we include the multiplets $({\bf 5, {10}})+({\bf \bar{10},
\bar{5}})$. We assume that below the GUT scale, only members of these multiplets
that remain light are the leptoquarks discused above.

Turning to unification with leptoquarks in this model, we first note that since
$U(1)_{B-L}$ and $SU(3)_c$ now emerge from two different
$SU(5)$'s and that there are more fermions in the fundamental representation of
the GUT group than the $SO(10)$ case,  the normalization of the low energy
couplings are totally different from the previous case. For instance, now
$I_Y=\sqrt{\frac{3}{13}} (\frac{Y}{2})$ and
$I_{BL}=\sqrt{\frac{3}{10}}(\frac{B-L}{2})$; furthermore,
$I_{a}=\sqrt{\frac{1}{2}}(I_{aL}+I_{aR})$ where $a$ denotes $SU(3)_c$ or $B-L$ 
generators. Because of this unification profile is now very different; however
with the same field content at $M_R$ as above, we find a unification scale of
$10^{9}$ GeV in the one loop approximation. In Table 1, we have  showed some
unification scenarios by varying the numbers of bidoublets.  We have one
$(2,2,4/3,3)+(2,2,-4/3.3^*)$ and one $(1,3,-2,1)+(1,3,2,1)$ and have used
$\alpha=1/127.9$, $sin^2\theta_w =0.2321$ and $\alpha_s =0.118$ at the $M_Z$
scale. From the table I, one can see that one can have one unified coupling at
the GUT scale using seven bidoublets. It is important to point out that this low
unification scale does not conflict with proton lifetime since proton is
absolutely stable in this model\cite{RNM}.  The essential reason for this is
that the vectorlike quarks and the light quarks do not mix with each other due
to the choice of the gauge group and Higgs representations.   

 We find it interesting that CDF collaboration, in their analysis of the
dilepton channel, have reported \cite{fabe} 4 events containing at least one 
$\tau$ (3 of the events have at least one b quark). Their estimated
background from non-$t\bar{t}$ productions is 2 events. Using their
measured cross-sections of 7.5 pb (from DIL, SVX and SLT channels),
and their $\tau$ channel acceptance of 0.12$\%$, only one such event
is expected from the $t\bar{t}$ productions in the 110 pb $^-1$ data.
It is entirely possible that this event is due to statistical
fluctuation, or due to charged Higgs production, however such excess
events are expected from the leptoquark or leptoquarkino
pair productions. One leptoquark decays to a $\tau$ and the top
quark, and the other to a e or a $\tau$. A leptoquark (or
leptoquarkino) pair production will have a branching ratio of about
0.75 for the dilepton channel with a $\tau$ in our model (compared to
about 0.05 in $t\bar{t}$ productions). This will give an acceptance
(branching ratio $\times$ detection efficiency) of about 1.8 $\%$.
Thus leptoquark (or leptoquarkino) pair production cross-section of
about 0.5 pb will give rise to such  an excess event. As can be seen
from fig 1, this corresponds to produced leptoquarkino mass about 250
GeV (or a leptoquark  of somewhat lower mass).          

In conclusion, we have suggested an alternative leptoquark  interpretation of
the HERA high $Q^2$ anomaly within an extension of the MSSM in such a way that
R-parity conservation is maintained  automatically. The superpartner of the
leptoquark, the leptoquarkino in this  model leads to many interesting and
testable predictions in present and future hadronic as well as in the future
e-gamma colliders. At Tevatron, the pair productions of these leptoquarkinos
give rise to very distinctive signal, namely opposite sign dilepton events of
the type ee, e$\tau$, $\tau\tau$ accompanied by dijets plus missing energy,
but no events of $\mu\mu$ type. If the leptoquarkino is not too much heavier
than the ``HERA leptoquark", then these events will  be observable in the
present or upgraded Tevatron.  It is possible  that one excess event in the
dilepton channel containing a $\tau$, as reported by the CDF
collaboration, could be due to the production of a leptoquark or
leptoquarkino. Although clearly more data is needed before definitive
conclusion can be drawn.     
        
The work of B. D has been supported by a DOE grant no. DE-FG06-854ER-40224; the
work of R. N. M. has been supported by NSF grant no. PHY-9421386 and the work of
S. N. has been supported by DOE grant no. DE-FG02-94-ER40852.

\begin{table}
\begin{tabular}{|c|c|c|c|c|}  \hline
$n_{\rm bidoublet}$& $M_I(GeV)$ &$M_G(GeV)$ &$\alpha _{G_A}^{-1}$&$\alpha
_{G_B}^{-1}$ \\\hline 5&$10^3$&$10^{9.11}$&5.12&2.29\\\hline
6&$10^3$&$10^{8.84}$&4.05&3.46\\\hline
7&$10^{3.75}$&$10^{9.04}$&4.25&4.25\\\hline
\end{tabular}
\caption{The Grand unification scale, the intermediate scale and the 
 strength of the unified couplings at the grand unified scale are shown for
different numbers of bidoublets. $\alpha _{G_A}$ is the unified coupling for one
$SU(5)$ group and $\alpha _{G_B}$ is the unified coupling for the other
$SU(5)$ group}
\end{table}

\begin{figure}[htb]
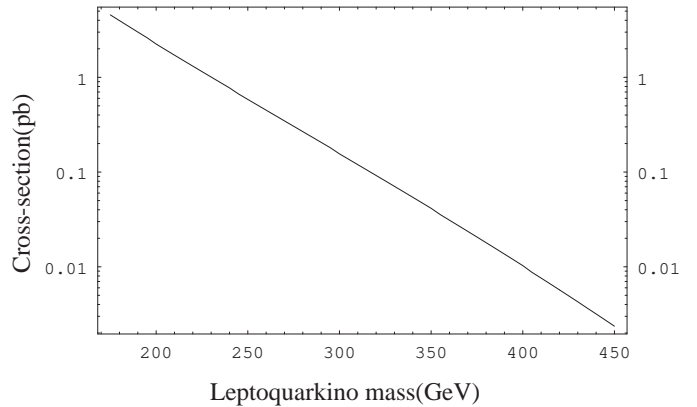

\vspace{1 cm}

\centerline{ \DESepsf(lqino.epsf width 9 cm) }
\smallskip
\caption {Cross section for the production of leptoquarkino pair at Tevatron
energy against the leptoquarkino mass.}

\end{figure}


\bigskip
                 


\begin{thebibliography}{99}
                 

\bibitem{H1} C. Adloff et al., H1 collaboration, DESY 97-024, hep-ex/9702012.

\bibitem{ZEUS} J. Breitweg et al., ZEUS collaboration, DESY 97-025,
hep-ex/9702015.

\bibitem{lq} For early discussions of leptoquarks, see R. N. Mohapatra, G. Segre
and L. Wolfenstein, Phys. Lett. {\bf 145B}, 433 (1984);  W. Buchmuller, R. Ruckl
and D. Wyler, Phys. Lett. {\bf 191B}, 442 (1987); S. Davidson, D. Bailey and B.
A. Campbell, Z. Phys. {\bf C61}, 613 (1994); M. Leurer, Phys. Rev. {\bf D 50},
536 (1994).  J. L. Hewett, Proc. 1990  Summer Study on High Energy Physics,
Snowmass, Co, ed. E.l. Berger,(world Scientific); J. Blumlein, hep-ph/9703287.

\bibitem{contact} K. S. Babu, C. Kolda, J. March-Russell and F. Wilczek,
hep-ph/9703299; M. C. Gozales-Garcia and M. Fabbrichesi, hep-ph/9703346; G.
Altarelli, J. Ellis, G. Giudice, S. Lola, M. Mangano, hep-ph/9703276;  V. Barger
et al., hep-ph/9703311; N. Di Bartolomeo and M. Fabbrichesi, hep-ph/9703375; A.
Nelson, hep-ph/9703379; W. Buchmuller and D. Wyler, hep-ph/9704317; K. Akama, K.
Katsuura and H. Terazawa, hep-ph/9704327.                  

\bibitem{godfrey} S. Godfrey, hep-ph/9704380.

\bibitem{squark} D. Choudhury and S. Raychaudhuri, hep-ph/9702392; G. Altarelli
et al., Ref. 4; K. S. Babu et al., Ref. 4; H. Dreiner and P. Morawitz,
hep-ph/9703279;  J. Kalinowski et al., hep-ph/9703288; T. Kon and T. Kobayashi,
hep-ph/9704221; J. Hewett and T. Rizzo, hep-ph/9703337; R. Barbieri, Z.
Berezhiani and A. Strumia, hep-ph/9704275.

\bibitem{haber} For a review see, H. Haber and G. Kane, Phys. Rep. {\bf 117}, 76
(1984).

\bibitem{vissani} For a review of the constraints on R-parity violating
couplings, see G. Bhattacharyya, {\it Supersymmetry96}, edited by R. N. 
Mohapatra and A. Rasin, North Holland (1997), p. 83.

\bibitem{mhkk} R. N. Mohapatra, Phys. Rev. {\bf D 34}, 3457 (1986); M. Hirsch,
H. Klapdor-Kleingrothaus and S. Kovalenko, Phys. Rev. Lett. {\bf 75}, 17 (1995).

\bibitem{moh} R. N. Mohapatra, \cite{mhkk}; A. Font, L. Ibanez and F. Quevedo,
Phys. Lett. {\bf B 228}, 79 (1989); S. Martin, Phys. Rev. {\bf D 46}, 2769
(1992).

\bibitem{pdg} R. M. Barnett et al. Phys. Rev. {\bf D 54}, 1 (1996).


\bibitem{cdf} F. Abe et al., Phys. Rev. Lett. {\bf 74}, 2626 (1995); for a
recent summary, see D. S. Kestelbaum, FERMILAB-CONF-97/ 016-E.

\bibitem{D0} S. Abachi et al., Phys. Rev. Lett. {\bf 74}, 2632 (1995); N.
Hadley, private communication for the most recent dilepton data.

\bibitem{talk} C. Grosso-Pilcher (CDF-collaboration), talk at the Vanderbuilt
meeting, May 15th,1997.

\bibitem{DN} M. Dine and A. Nelson, Phy. Rev. {\bf D47}, 1277 (1993); M. Dine,
A. Nelson and Y. Shirman, Phys. Rev. {\bf D51}, 1362 (1995); M. Dine, A. Nelson,
Y. Nir and Y. Shirman, Phys. Rev. {\bf D53}, 2658 (1996); M. Dine, Y. Nir and Y.
Shirman, preprint SCIPP-96-30, hep-ph/9607397.



\bibitem{don} M. Doncheski, S. Godfrey, Talk given at 1996 DPF / DPB Summer
Study on New Directions for High-Energy Physics (Snowmass 96), Snowmass, CO, 25
Jun - 12 Jul 1996, hep-ph/9612385 and references therein. 

\bibitem{RNM} R. N. Mohapatra, Phys. Lett. {\bf B379}, 115 (1996).

\bibitem{fabe} F. Abe et al (CDF collaboration), CDF/ANL/TOP/PUBLIC/4011,(1997). 

\end{thebibliography}
\end{document}